# Solitary beam propagation in a nonlinear optical resonator enables high-efficiency pulse compression and mode self-cleaning

Sheng Zhang[1†], Zongyuan Fu[1†], Bingbing Zhu[1†], Guangyu Fan[2], Shunjia Wang[1], Yudong Chen[1], Yaxin Liu[1], Andrius Baltuska[2], Chuanshan Tian[1], Zhensheng Tao[1*]

[1]*State Key Laboratory of Surface Physics and Department of Physics, Fudan University, Shanghai, China*

[2]*Institute of Photonics, TU Wien, Gusshausstrasse 27/387, Vienna, Austria*

[†]These authors contributed equally to this work.

*Corresponding authors: Dr. Zhensheng Tao, ZhenshengTao@fudan.edu.cn.

## Abstract

Generating intense ultrashort pulses with high-quality spatial modes is crucial for ultrafast and strong-field science. This can be accomplished by controlling propagation of femtosecond pulses under the influence of Kerr nonlinearity, and achieving stable propagation with high intensity. In this work, we propose that the generation of spatial solitons in periodic layered Kerr media can provide an optimum condition for supercontinuum generation and pulse compression using multiple thin plates. With both the experimental and theoretical investigations, we successfully identify these solitary modes and reveal a universal relationship between the beam size and the critical nonlinear phase. Space-time coupling is shown to strongly influence the spectral, spatial and temporal profiles of femtosecond pulses. Taking advantages of the unique characters of these solitary modes, we demonstrate single-stage supercontinuum generation and compression of femtosecond pulses from initially 170 fs down to 22 fs with an efficiency ~90%. We also provide evidence of efficient mode self-cleaning which suggests rich spatial-temporal self-organization processes of laser beams in a nonlinear resonator.

Generating high-energy, ultrashort pulses with high-quality spatial modes is essential for numerous applications in ultrafast optics and strong-field physics. A route to shorter pulse durations is nonlinear pulse compression, which relies on supercontinuum generation (SCG) enabled by self-phase modulation (SPM) in a $\chi^{(3)}$ nonlinear medium, in combination with a negative dispersive delay line that compensates for the spectral chirp [1]. In past years, great effort has been made to in manipulating the characteristics of femtosecond pulses and engineering the nonlinear medium for SPM, in order to achieve sustainable light-matter interaction for generating ultrabroad-band SCG. For instance, with the spatial mode confined by gas-filled hollow-core fibers (HCFs), femtosecond pulses can interact with gas medium for several meters, which enables compression of pulses with high energy [2], high average power [3], and can provide high compression ratios [4]. However, the SCG technique based on HCFs is usually sensitive to alignment, hard to maintain, and the fiber-coupling efficiency is generally <60%. More recently, gas-filled multi-pass cell has further achieved light-gas interaction for hundreds meter long with an efficiency of ~90% [5–7]. But, owing to the large number of reflections on the cavity concave mirrors, they require state-of-the-art coating with low group-delay dispersion (GDD) and high reflection, which is costly and technically challenging. So far, that all these efforts have been directed towards realizing long-distance propagation of intense femtosecond pulses under the influence of Kerr nonlinearity with stable profiles in both space and time, which, in fact, can be accomplished by soliton formation [8].

Compared to gas media, SCG in condensed media is advantageous, because it is simple, flexible and robust. The nonlinear refractive index $n_2$ of condensed media is typically 1000 times larger, which could lead to compact device footprint. Nevertheless, the propagation of strong femtosecond pulses in a bulk medium is more challenging, also because of the high nonlinearity. The self-focusing effect can result in catastrophic beam collapse and optical damage, when the power is higher than the critical power $P_{cr}$ [8]. Recently, the progress in SCG with multiple thin plates of Kerr media has attracted great attention [9–14]. By placing the beam self-focusing in the free space between material plates, this technique elaborately circumvents the optical breakdown, allowing substantial enhancement of nonlinear Kerr interaction [10,12]. Although generation of few-cycle pulses has been demonstrated with a two-stage compressor [13], the existing implementations of multiplate SCG and pulse compression have several limitations, because the beam propagation is still not well controlled. First, complicated space-time coupling leads to strong conical emission, which can cause energy loss >40% [12]. Second, because of the space-time-coupling-induced higher-order dispersion, the dispersion compensation requires custom-designed chirped mirrors or pulse shapers in some cases [12,13], and strong pedestals can be occasionally observed in the compressed pulses [13]. Finally, the plate thickness and their positions have been empirically determined for the broadest spectrum, making it difficult to be systematically studied and repeated.

Here, we want to point out that the optimum condition for SCG in layers of Kerr media is to form discrete solitons, which can sustainably propagate and interact with the medium under high intensity. Discrete solitons are localized wave-packets propagating with a stable structure in nonlinear periodic structures, thanks to the balance between diffraction and material nonlinearity [8,15]. Fundamentally, understanding the formation of solitons is important, because they represent stable solutions for the cubic nonlinear Schrödinger equations (NLSE) [16,17], which is related to many branches of physics [18,19]. Theory has predicted that it is possible to have solitary spatial modes in periodic layered Kerr media (PLKM) [20–22]. Experimentally, repetitive variation of the beam profiles for $P \approx 6\ P_{cr}$ has been observed without spatial beam collapse [23]. Ultimately, such periodic propagation of femtosecond pulses can be considered as two-dimensional transverse [(2+1)D] discrete spatial solitons, with a repetitive spatial mode on the material layers [15,21]. The stability of these solitons, however, could be strongly influenced by space-time coupling during the propagation. To date, no experimental studies on the influence of spatiotemporal propagation exist. Neither have any shown how to identify these solitary modes, or studied their potential applications for SCG and pulse compression.

In this Letter, we experimentally characterize the space-time coupled propagation of femtosecond pulses with a peak power reaching GW in a nonlinear resonator comprising PLKM, and successfully identify the formation of discrete solitary modes under a range of conditions. By comparing the experimental and theoretical results, we reveal a universal relationship between the characteristic beam size and the critical nonlinear phase of the solitary modes, under which the propagation of light wave-packets is localized in both space and time. We also show that the formation and breakdown of solitons is manifested by the correlated variations in spectral, spatial and temporal profiles of femtosecond pulses. In practice, we demonstrate two applications of these solitary modes – First, taking advantage of the localized solitary modes, we can successfully suppress the spatial loss and higher-order dispersion, and achieve ~8-fold pulse compression with ~90% efficiency in a single-stage compressor. This result demonstrates the unique advantages of the solitary modes over the empirically implemented multiplate

SCG [9,10,13]. Second, spatial mode-self-cleaning with high efficiency is demonstrated when the beam propagation is on resonance, which can be attributed to the spatial self-organizing effect in a nonlinear resonator.

The repetitive propagation of a high-intensity laser beam in PLKM can be regarded as a cavity resonator with intensity-dependent non-spherical (Kerr-lens) mirrors (see Fig. 1(a)). Because of the complexity of the spatiotemporal effects introduced by Kerr nonlinearity [8], the NLSE simulation relies heavily on numerical analysis, which complicates understanding of the fundamental physical processes. Here, we first resort to the Fresnel-Kirchhoff diffraction (FKD) integral to identify the self-consistent solitary modes. Assuming the normalized amplitude of the incident optical field as $U_1$, we derive that the amplitude, after propagating through a unit of the resonator and right before the next period, is given by

$$U_2(\rho) = -2\pi j e^{j\pi\rho^2} \int_0^\infty U_1(\rho') e^{jb|U_1(\rho')|^2} \cdot e^{j\pi\rho'^2} J_0(2\pi\rho'\rho) \rho' d\rho', \tag{1}$$

where $\rho$ and $\rho'$ are the radial coordinates rescaled by $\sqrt{\lambda L}$ and $J_0$ the zeroth-order Bessel function. Here, one period of the resonator contains a layer of Kerr medium with a thickness of $l$ and the following layer of free space by length $L$. In Eq. (1), $b$ represents the nonlinear phase given by $b = \frac{2\pi}{\lambda} n_2 l I_0$, where $I_0$ is the field intensity. The Fox-Li iteration is then used to numerically find the solitary modes [24], the nonlinear phase associated with which is defined as the critical nonlinear phase $b_c$. As shown in Fig. 1(b), the characteristic properties of the solitary modes are determined by the beam radius ($w$) on the layers, the resonator length ($L$) and the critical nonlinear phase ($b_c$). Here, we define a Fresnel-number-like beam radius squared $w^2/\lambda L$, for convenience of discussion.

In our experiments, we employed a Yb:KGW amplifier laser system with a pulse duration of 170 fs at $\lambda$=1030 nm. Transform-limited femtosecond pulses with p polarization are focused in to a PLKM resonator (see Supplementary Materials (SM) for the experimental setup). The PLKM is composed of polycrystalline $Al_2O_3$ thin plates as the Kerr medium, placed at the Brewster angle to minimize the loss on each plate <0.5%. The nominal thickness of the plates is fixed to be 0.4 mm (see SM) and the distance between the neighboring plates equals to the resonator length $L$. By adjusting the incident pulse energy, the resonant modes are experimentally identified. The evolution of spatial beam profiles is monitored with a 4-$f$ imaging setup, while the temporal intensity profiles are measured by second-harmonic-generation frequency-resolved optical gating (SHG-FROG) [25].

First, we present findings on the key features of the resonator stability and soliton formation. In the experiments, we observe a reduction of the far-field beam size and an improvement of the spatial mode quality (see SM), as the incident pulse energy ($E_{in}$) approaches from low to a critical value under a specific resonator length $L$ (e.g. $L$=50.8 mm in the inset of Fig. 1(b)). When the pulse energy is low and the self-focusing in the media is weak, the beam propagation is dominated by diffraction, leading to a diverging beam size on the successive layers ("*unstable diffractive region*" in Fig. 1(b)). On the contrary, when the incident beam reaches the critical nonlinear phase ($b_c$) and produces sufficient self-focusing to appropriately balance the diffraction ($E_{in}$=260 μJ for $L$=50.8 mm, corresponding to a field intensity of $5.0\times10^{12}$ W/cm$^2$), the laser beam can repetitively propagate through the PLKM with a well-confined beam size. Indeed, with the 4-$f$ imaging measurements, we observe that the beam radius approaches a stable value on the material layers after the self-adjustment in the first few layers under the resonant condition (Fig. 1(c)), which indicates the formation of discrete spatial solitons [21]. We summarize the experimental results under different $L$s in Fig. 1(b), which exhibits excellent agreement with the FKD model. We note that the excellent agreement between our results and the FKD model indicates that the formation of spatial solitons is fundamentally caused by the balance between the diffraction and nonlinear self-focusing effects. Meanwhile, since the temporal profiles are not considered in the FKD model, such agreement also suggests that the femtosecond pulses should have a stable temporal structure under the spatial soliton modes throughout the propagation. This is confirmed by the direct measurement of the temporal profiles, as shown in Fig. 1(d). As a result, we can define these solitary modes as temporally confined (2+1)D spatial solitons.

As shown in the inset of Fig. 1b, the beam size does not immediately diverge when $E_{in}$ rises above the critical value. According to the 1D FKD model, this corresponds to a region, where the beam size oscillates as it propagates through the PLKM ("*quasi-stable oscillatory region*", see SM). The upper boundary of this region is approximately $P = 2P_{thr}$, where $P_{thr}$ is given by $P_{thr} = P_{cr}\gamma n_0$ [22]. The geometrical factor $\gamma$ is given by $\gamma = \frac{L}{l}$. However, we find that the width of this region is generally narrower in the experiments than the FKD results. This could be attributed to the temporal pulse splitting above the resonance (see below), which is not considered in the FKD model. Beyond $2P_{thr}$, the strong Kerr lens breaks the balance between the self-focusing and diffraction, and the beam size grows out of limit within few periods of propagation ("*unstable dissipative region*", Fig. 1b).

In the regions beyond the resonance, the space-time coupling breaks the solitary modes and strongly influences the spectral, spatial and temporal profiles (see Fig. 1a). First, we find that the bandwidth ceases to increase almost immediately when $b>b_c$, as shown in Fig. 2a.

Correspondingly, we observe temporal pulse splitting and asymmetric pulse profiles (Fig. 2c), accompanied by the strong enhancement of conical emission. These effects can be understood as the correlated effects resulting from the spatial cavity-mode selection for different pulse intensities in time. When the peak intensity is higher than the critical value, the pulse temporal center mismatches with the cavity resonance and experiences fast divergence due to the strong Kerr lensing. This breaks the solitary modes and results in splitting of pulses in time and conical emission (Fig. 1(a)). This explains the widely observed strong conical emission in many previous experiments [10,12,13]. The saturation of spectral bandwidth, on the other hand, is caused by the SPM process on a temporally split and positively chirped pulse [26]. These correlated effects can be observed for different resonator lengths, indicating the universality of these effects. Meanwhile, they can be well captured by the 2D NLSE simulations with an accuracy of $E_{in}$ within 10% of the experimental values (see SM).

In Fig. 2(d) and (e), we plot the time-frequency analysis of the output pulses under different conditions (see SM). In the dissipative region, significant amount of nonlinear chirp can be observed in the long-wavelength side of the spectrum (Fig. 2(e)). In stark contrast, majority of the optical energy is linearly chirped under the solitary mode (Fig. 2(d)). We note that, since the pulses have passed through the same amount of Kerr medium in both cases, we can exclude that such difference is caused by the higher-order dispersion of materials or carried by the input pulses. On the other hand, when the femtosecond pulses are split and asymmetric in time, nonlinear frequency chirp can be produced by the space-time coupling and SPM process [27]. Indeed, as we have shown in Fig. 1(d), the complex pulse profiles have already been generated through few material layers in the dissipative region. The higher-order dispersion here has significant effects on pulse compression. Even with an appropriate compensation of negative GDD, strong pedestals spanning ~100 fs in time can still be clearly observed for the compressed pulse in Fig. 2(e), which contributes to ~24% of the overall pulse energy (SM). In contrast, the pedestals can be well suppressed when the pulse propagation is under the solitary mode (see Fig. 3(a)).

The generation of GW, temporally confined spatial solitons in this work can significantly improve the SCG and pulse compression. Here, we summarize three advantages. First of all, as evidenced in Fig. 1(c) and Fig. 1(d), the solitary mode can maintain the high intensity throughout the propagation in PLKMs, supporting sustainable nonlinear light-matter interaction over many layers of the Kerr media. As an example, we send 260 μJ, 170 fs pulses through a 15-layer PLKM with $L$ = 50.8 mm under the resonant condition. The spectrum of the output pulse is significantly broadened, corresponding to a transform-limited (TL) pulse duration of 22 fs. The chirp of the output pulses is then compensated by a set of chirped mirrors which supplies a negative GDD of -1200 fs$^2$ over 850-1200 nm. The duration of the compressed pulses from this single-stage compressor is close to the TL pulse (see Fig. 3(a)). The experimental and reconstructed FROG traces are shown in Fig. 3(b) and (c), respectively. We note that our result represents the greatest number of layers implemented in a single-stage multiplate SCG experiment, and the SCG bandwidth is about 50% broader than the previous work under the similar conditions [13]. Secondly, the single-peaked temporal profile under the solitary modes (Fig. 2(f)) also avoids generating higher-order dispersion, which circumvents the usage of custom-designed chirped mirrors or pulse shapers [12,13]. This is evidenced by the fact that the clean pulse compression (Fig. 3(a)) is achieved by only compensating the second-order dispersion. Last but foremost, the spatial loss induced by conical emission is strongly suppressed by the solitary propagation. As shown in Fig. 3(d), the conical radiation only contributes <10% of the total output energy when the propagation is on resonance, and this contribution can increase to ~35% in the dissipative region (Fig. 3(e)). Overall, by combining the broad SCG spectrum and suppressed loss from space and time, we achieve ~5 times increment in the peak power in a single-stage compressor, as shown in Fig. 3(a), which represents the largest increment of pulse peak power from a single-stage compressor with multiple layers of Kerr media (see SM).

Finally, we show the spatial mode self-cleaning from the nonlinear PLKM resonator. The improvement of the output spatial mode, in terms of the circularity and the intensity profile, can already be observed after the fundamental laser beam passing through the PLKM (see SM). To further investigate the mode-self-cleaning effect, we introduce substantial perturbance on the beam profile, by inserting a cylindrical beam blocker with a diameter of 0.8mm into the laser beam, which has a full-width-of-half-maximum (FWHM) size of ~3 mm (Fig. 4(a)). The spatially modulated laser beam is then focused into a resonator with $b_c$=0.5, $L$=25.4 mm, consisting of 20 layers. As shown in Fig. 4(b-d), by matching with the cavity resonance, the output mode is significantly cleaned and transforms to the resonator solitary mode. In Fig. 4(e), we plot the filtering efficiency as a function of the blocker transverse position $\Delta x$, with larger $\Delta x$ inducing greater spatial modulation (Fig. 4(a)). The PLKM resonator can support a high filtering efficiency (>85%) across a large range of spatial modulation, in direct contrast to an ideal linear spatial filter (see SM). This result suggests that the solitary modes here represents attractors of the (2+1)D cubic NLSE equation [18]. There must be an efficient pathway, in which the laser energy in the higher-order spatial modes can be transferred to the solitary modes through the repetitive Kerr interactions. This is consistent with the spatial self-organization of laser beams, previously observed under the nonlinear interactions in filamentation [28], self-focusing collapse [29] and multimode fibers [30,31].

In summary, we experimentally investigate the spatiotemporal propagation of strong femtosecond pulses in PLKM resonators and reveal its influence on the stability of optical solitons. Taking advantages of the unique characters of these solitary modes, we demonstrate high-efficiency SCG, pulse compression and spatial mode-self-cleaning. These results are relevant to a wide range of applications, such as Kerr-lens mode locking [32], ultrashort-pulse generation [10,12,13] and high-energy wavelength scaling [11]. Moreover, the results here illustrate the general features of the space-time coupling of solitons under periodically modulated Kerr nonlinearity, which may reinforce the theory and help understanding soliton formation under similar periodic "potentials" in many other fields of nonlinear optics, including waveguide arrays [33,34], periodic refractive-index gratings [35,36] and photonic crystal fibers [37], as well as in condensed matter physics [38,39] and in biology [40].

**Acknowledgement**

We thank Aleksei Zheltikov, Huailiang Xu, Andy Kung and Ming-Chang Chen for helpful discussions. The experimental studies were performed at Fudan. We gratefully acknowledge the financial support from the National Natural Science Foundation of China (Grant No. 11874121) and the Shanghai Municipal Science and Technology Basic Research Project (Grant No. 19JC1410900). CS. T. acknowledges support from the National Natural Science Foundation of China (No. 11874123) and the National Key Research and Development Program of China (No. 2016YFA0300902). Z. T. thanks the support from the Alexander-von-Humboldt foundation.

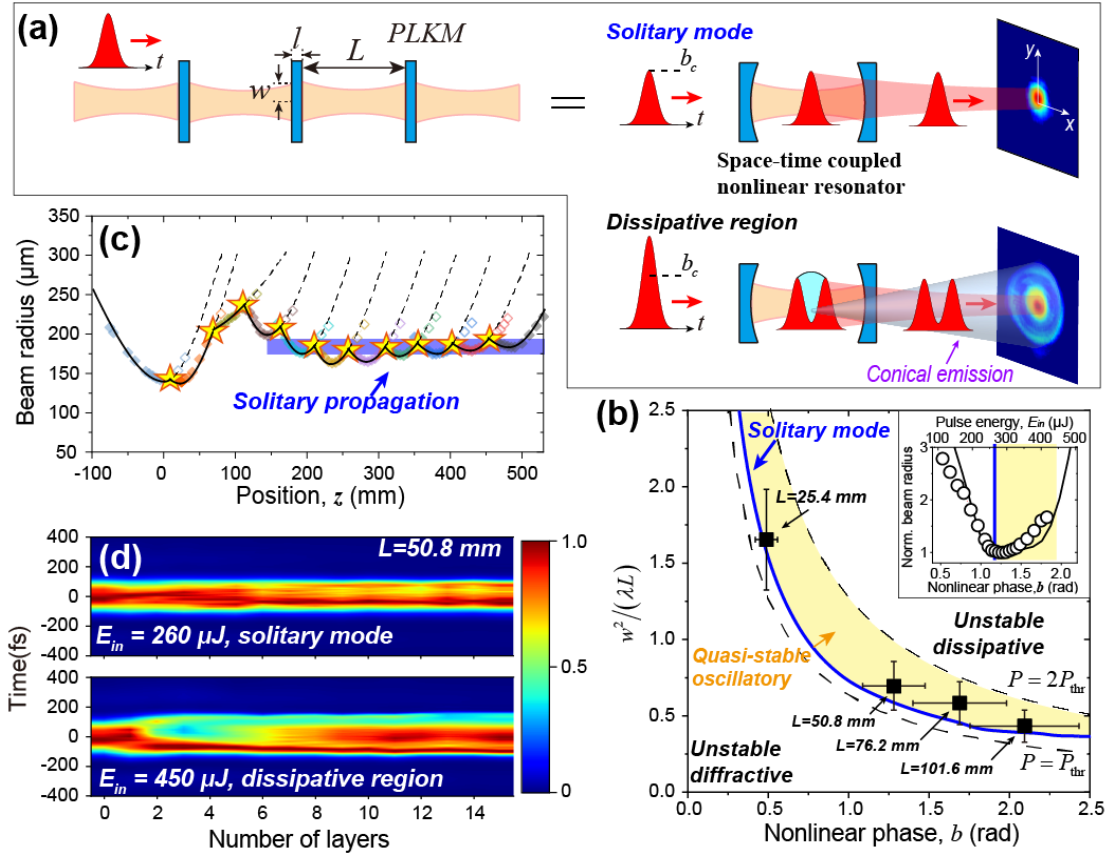

**Figure 1.** (a) Illustration of the repetitive self-focusing and diverging pattern in PLKM, which is equivalent to a space-time coupled nonlinear optical resonator. The space-time coupled propagation of femtosecond pulses for the solitary modes and in the dissipative region is presented. When the propagation is in the dissipative region, strong conical emission is contributed by the optical energy at the temporal center of the pulse, leading to temporal pulse splitting. (b) Universal relationship of the normalized beam radius squared $w^2/\lambda L$ as a function of the nonlinear phase $b$ for the resonator solitary modes (solid blue line). The dashed lines represent the boundaries for $P_{thr}<P<2P_{thr}$. The symbols are the experimental results with different resonator lengths. The yellow shaded area labels the "quasi-stable oscillatory region". Inset: Variation of the far-field beam radius as a function of applied nonlinear phase and incident pulse energy for $L$=50.8 mm, $b_c$ = 1.2. (c) Experimentally measured beam radii through the propagation in the PLKM with $L$=50.8 mm, $b_c$ = 1.2 under the resonant condition. The beam radius on each layer is labeled by the star symbols. (d) Evolution of temporal intensity profiles as the pulses propagate through the PLKM under the solitary mode ($E_{in}$ =260 μJ) and in the dissipative region ($E_{in}$ =450 μJ).

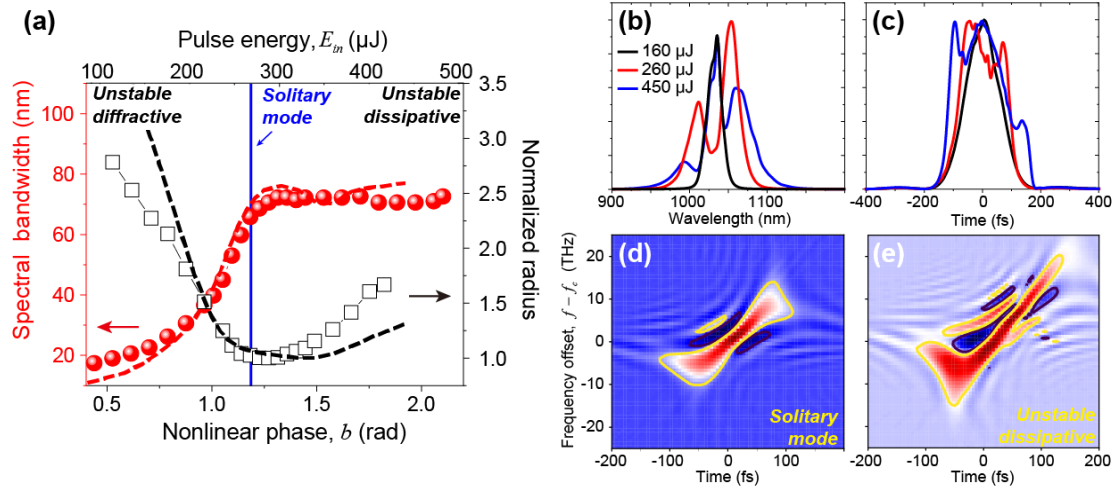

**Figure 2.** (a) The spectral bandwidth and normalized beam radius as a function of the nonlinear phase and pulse energy for the PLKM resonator with $L$=50.8 mm, 15 layers. The dashed lines are the results of the NLSE simulations. (b-c) The far-field spectral and temporal profiles in different stability regions. (d) Time-frequency analysis (Wigner plots) of the output pulses under the solitary mode. $f_c$ is the central frequency. The contours label the $1/e^2$ intensity in the figure. (e) Same as (d) for the output pulses in the dissipative region.

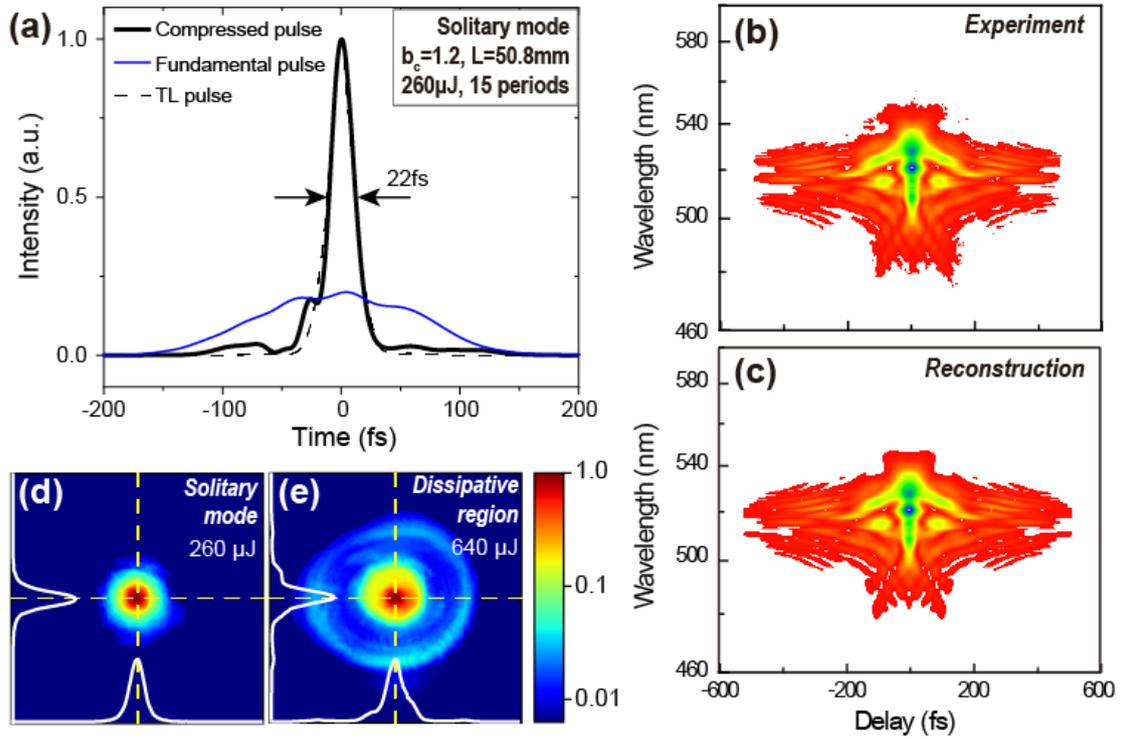

**Figure 3.** (a) The temporal intensity profiles of the compressed pulse (solid black) and the incident pulse (solid blue). The dashed black line is the TL pulse. (b) and (c) The measured and reconstructed FROG traces. (d) The spatial mode of output beam on resonance ($E_{in}$ = 260 μJ). The intensity is plotted in log-scale to highlight the conical emission. The white lines display the $x$ and $y$ transverse profiles of the beam. (e) Same as (d) with $E_{in}$ = 640 μJ in the dissipative region.

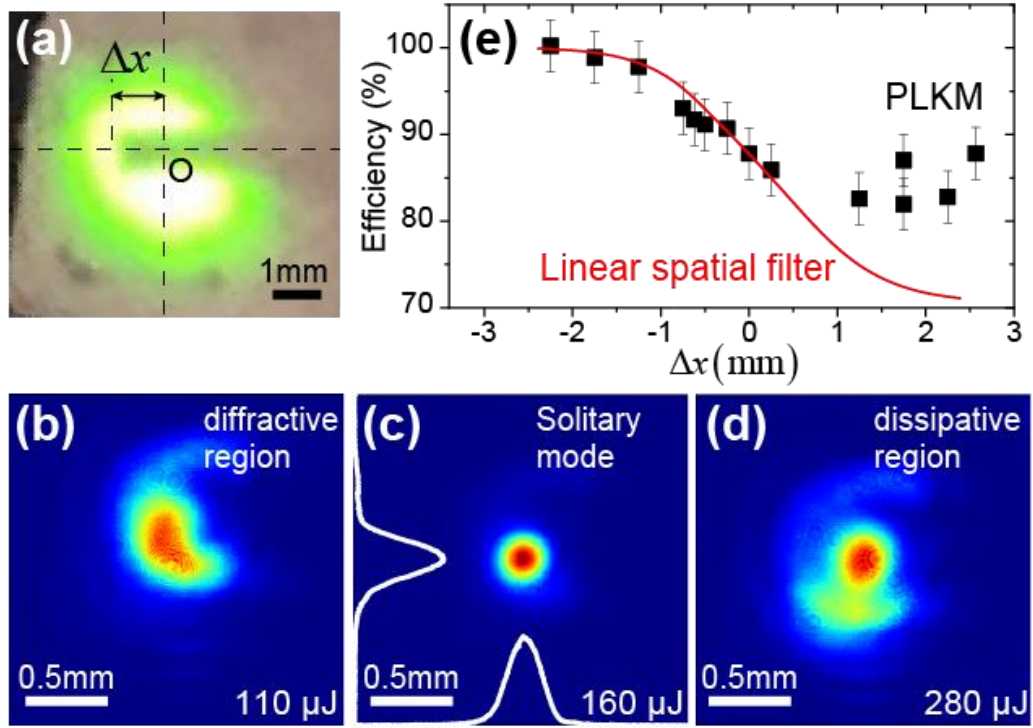

**Figure 4.** (a) Camera picture of the modulated laser beam after the beam blocker for Δx≈1.25mm. O is the center of the beam profile and Δx is the transverse blocker position. (b-d) The output beam profiles measured in the far field, for the spatially modulated input beam shown in (a), in the diffractive region, at the solitary mode and in the dissipative region, respectively. The lineouts in (c) represents the horizontal and vertical beam profiles at the solitary mode. (e) The filtering efficiency as a function of the blocker position (Δx) for the PLKM resonator and a linear spatial filter.